\pdfoutput=1
\documentclass[
    aps,prl,twocolumn,
    reprint,
    superscriptaddress,
    floatfix,
    amssymb,
    longbibliography
]{revtex4-1}

\usepackage{amsmath}
\usepackage{amsfonts}
\usepackage{graphicx}
\usepackage{bm}
\usepackage{dcolumn}
\usepackage{lipsum}
\usepackage{dsfont}
\usepackage{nicefrac}
\usepackage[normalem]{ulem}
\usepackage[dvipsnames]{xcolor}
\usepackage{xcolor}
\usepackage{comment}

\definecolor{linkColor}{rgb}{0,0.3,0.7}
\usepackage{hyperref}
\hypersetup{colorlinks=true,
            allcolors=linkColor,
            pdfborder={0 0 0},
            pdfencoding = auto
            }

\begin{document}

\title{Stochastic bubble dynamics in phase-separated scalar active matter}

\author{Mingqi Yan}
\affiliation{Arnold Sommerfeld Center for Theoretical Physics and Center for NanoScience, Department of Physics, Ludwig-Maximilians-Universität München,
Theresienstraße 37, D-80333 München, Germany}
\affiliation{
Institut für Theoretische Physik, Georg-August-Universität Göttingen, Friedrich-Hund-Platz 1, D-37077 Göttingen, Germany
}
\affiliation{%
Institut für Dynamik komplexer Systeme, Georg-August-Universität Göttingen, Friedrich-Hund-Platz 1, D-37077 Göttingen, Germany
}
\affiliation{
Max Planck School Matter to Life, Hofgartenstraße 8, D-80539 München, Germany
}
\author{Erwin Frey}
\email{frey@lmu.de}
\affiliation{Arnold Sommerfeld Center for Theoretical Physics and Center for NanoScience, Department of Physics, Ludwig-Maximilians-Universität München,
Theresienstraße 37, D-80333 München, Germany}
\affiliation{
Max Planck School Matter to Life, Hofgartenstraße 8, D-80539 München, Germany
}
\author{Marcus Müller}
\email{marcus.mueller@uni-goettingen.de}

\affiliation{
Institut für Theoretische Physik, Georg-August-Universität Göttingen, Friedrich-Hund-Platz 1, D-37077 Göttingen, Germany
}%
\affiliation{
Max Planck School Matter to Life, Hofgartenstraße 8, D-80539 München, Germany
}%
\author{Stefan Klumpp}
\email{stefan.klumpp@phys.uni-goettingen.de}
\affiliation{%
Institut für Dynamik komplexer Systeme, Georg-August-Universität Göttingen, Friedrich-Hund-Platz 1, D-37077 Göttingen, Germany
}%
\affiliation{
Max Planck School Matter to Life, Hofgartenstraße 8, D-80539 München, Germany
}


\begin{abstract}
In ABP systems, phase separation is accompanied by the emergence of vapor bubbles within liquid domains. Using large-scale particle-based simulations, we study the stochastic dynamics of these bubbles and find that most nucleate, grow, and dissolve within liquid domains. We show that their area dynamics can be described by a Langevin equation with a constant negative drift and noise proportional to the perimeter, fully characterizing bubble area and lifetime statistics. Additionally, we develop a lattice gas model that captures the morphological properties, including the decrease in bubble asphericity with increasing area. These findings provide new insights into phase separation in active matter and highlight limitations in current continuum theories.
\end{abstract}


\maketitle

Steady states in nonequilibrium systems exhibit distinctive features absent in thermal equilibrium systems, such as complex spatial structures and dynamic behaviors.
This includes active matter systems~\cite{Bowick.2022}, where interacting particles continuously convert energy into motion and forces, and nonequilibrium biological processes~\cite{Burkart.2022}, which are sustained by chemical reactions that break detailed balance.
Among the simplest representative of such nonequilibrium systems are those where the dynamics can be described using scalar fields. 
These fields might represent quantities like the density of active particles~\cite{2015AnnualReviewsCates} or the concentrations of a few key chemical species~\cite{Frey_Brauns.2022}. 

Here, we focus on systems of Active Brownian Particles (ABPs) with repulsive interactions, which have been shown to undergo phase separation into dense liquid domains and vapor, as evidenced by both computer simulations~\cite{Fily2012,2013FirstMSDPRL} and experiments~\cite{2013ABPExperiments}.
Notably, this class of phase separation in active matter systems---known as motility-induced phase separation---occurs in the absence of attractive interactions. 
This phenomenon was predicted on the basis of a scalar field theory referred to as Active Model B~\cite{2013PRL_FirstActiveModelB}. 
It builds on the Cahn-Hilliard model~\cite{1958_Cahn_Hilliard}, a classical framework for phase-separation dynamics of thermal equilibrium systems, also known as Model B~\cite{1977_Hohenberg_Halperin}. 
Active Model B is still grounded in symmetry arguments and leading-order gradient expansions, but includes nonequilibrium particle currents that cannot be expressed as gradients of a free-energy functional.
Interestingly, numerical simulations discovered that the liquid domains contain a population of mesoscopic vapor bubbles~\cite{2014ABPDimensionality,Marchetti2018}, a phenomenon that was not present in scalar field theories~\cite{2013PRL_FirstActiveModelB}. 
To account for these numerical observations, Active Model B+~\cite{2018PRXABMP} was introduced, incorporating both irrotational and rotational components of the particle current in the spirit of a Helmholtz decomposition.
This extended model reveals a parameter regime with reverse Ostwald ripening, resulting in microphase separation 
into liquid domains where bubbles form, migrate to the boundaries, and exit into the surrounding vapor \cite{2018PRXABMP,Nardini2024}.

Despite these theoretical advances, the dynamics of the bubbles in the liquid domains remain poorly understood. 
The present study addresses this gap by investigating their dynamics using particle-based simulations, with a particular focus on how the area and shape of the bubbles evolve with time. 
By complementing these numerical studies with an effective Langevin equation for the bubble area, we provide new insights into bubble dynamics. 
Most bubbles nucleate, grow, and dissolve within the liquid domains, exhibiting significant fluctuations in their area. 
The positional dynamics of the bubbles follow a random walk, characterized by a self-diffusion coefficient inversely proportional to the bubble radius and lacking any net flux toward the periphery. 
The stochastic dynamics of bubble area is governed by a process with a constant negative drift and a diffusion coefficient proportional to the bubble perimeter. 
This leads to a steady-state distribution of bubble areas described by a generalized Gibbs distribution, with bubble lifetimes scaling proportionally to their area. 
Morphological and dynamical properties of bubbles are effectively captured by a lattice gas model. 
Furthermore, the stochastic dynamics of bubble area represent a broader class of stochastic processes with sub-demographic noise, possessing intriguing properties in their own right.

\textit{Particle-based simulations.}
We consider a particle-based system of ABPs confined to two spatial dimensions governed by the Langevin equations~\cite{2013FirstMSDPRL,2014ABPDimensionality,2015PRLNegative}
\begin{equation}
\begin{gathered}
    \dot{\boldsymbol{x}}_i
    =
    v_0 \hat{\boldsymbol{e}}_i-  \nabla_i U\left(\left\{\boldsymbol{x}_i\right\}\right)/\gamma+\sqrt{2 D_{\mathrm{T}}} \, \boldsymbol{\eta}^T_i(t)  
    \, ,
    \\
    \dot{\theta}_i
    =
    \sqrt{2 D_{\mathrm{R}}} \, \eta_i^R 
    \, .
\end{gathered}
\end{equation}
Here ${\hat{\boldsymbol{e}}_i = (\cos \theta_i, \sin \theta_i)}$ denotes the velocity direction, $v_0$ the constant particle speed, and ${\gamma = k_{\mathrm{B}}T/D_{\mathrm{T}}}$ the translational friction coefficient. 
The spherical particles (with diameter $\sigma$) are assumed to interact through a repulsive, pairwise additive Weeks–Chandler–Andersen (WCA) potential ${U\left(\left\{\boldsymbol{x}_i\right\}\right) = \sum_{i<j} u\left(\left|\boldsymbol{x}_j-\boldsymbol{x}_i\right|\right)}$ 
with ${u(r) = 4 k_{\mathrm{B}}T  \big( (\sigma/r)^{12}-(\sigma/r)^{6}+1/4 \big)}$ for ${r < 2^{1 / 6} \sigma}$, above which ${u = 0}$ \cite{1971_WCA}.
For both the translational and rotational motion, the noise is assumed to be Gaussian white noise with unit variance.  The noise amplitudes are given in terms of the translational and rotational diffusion constants, which in the low-Reynolds-number regime are related by ${D_{\mathrm{R}}=3 D_{\mathrm{T}}/\sigma^2}$ ~\cite{guazzelli_morris_pic_2011,2023AnnualReview}.

We use the particle diameter  $\sigma$ as the unit of length, $k_\mathrm{B}T$ as the unit of energy, and the diffusion time ${\tau:=\sigma^2 / D_\mathrm{T}}$ as the unit of time. 
The Péclet number ${\mathrm{Pe} := v_0 \tau/\sigma}$ quantifies the ratio of the timescale for a particle to travel its own diameter by diffusion to the timescale for ballistic motion. 
Simulations were carried out using the Euler-Maruyama scheme with a timestep ${\delta t = 10^{-5}}\ \tau$ for ${N = 1024^2}$ particles within a square box of side length ${L=1085\ \sigma}$ with periodic boundary conditions.
The large system size, with over one million particles, is an essential prerequisite for the reliable observation of phase separation and bubble formation in the liquids.
The simulation was run for ${T = 5000  \, \tau}$ to obtain robust statistical results. To ensure effective phase separation, we chose a high Péclet number of $ {\mathrm{Pe} = 140}$ and a high volume fraction of $0.7$.

\begin{figure}[!t]
\includegraphics[width=\columnwidth]{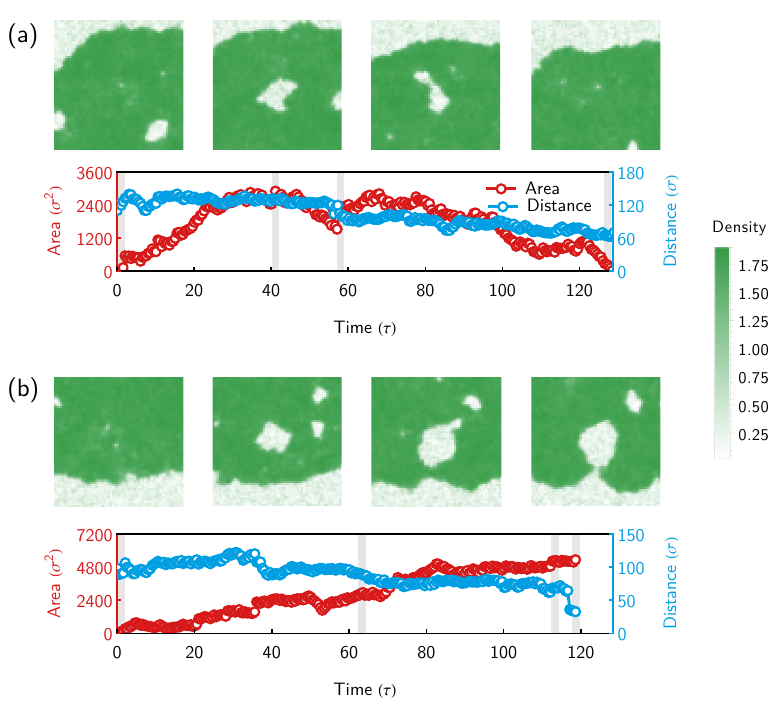}
\caption{
\label{fig:simulations_illustration}
Vapor bubbles in the liquid domain of a phase-separated ABP system may either form and dissolve within liquid domains (a) or escape through the domain boundary (b). 
The time evolution of representative bubbles (arrows) is illustrated as a sequence of simulation snapshots with particle density (color scale) given in units of the initial average value 17.0489 $\xi^{-2}$. 
Their dynamics is quantified by changes in area (red) and distance from the domain boundary (blue); grey vertical bars mark the time points of the simulation snapshots. See Sec.~3 of \cite{supplemental_material} for details.
}
\end{figure}

\textit{Stochastic bubble dynamics.} 
In our particle-based simulations, we observed motility-induced phase separation into liquid and vapor domains within a few $\tau$. 
As the size of the liquid domains increased, bubbles were observed to form around $10\ \tau$; cf. Video~1~\cite{supplemental_material}.
These bubbles emerge randomly in both space and time, triggered by a local decrease in liquid density, which is then followed by an expansion in bubble area [Fig.~\ref{fig:simulations_illustration}].
To analyze the bubble dynamics, we measured three key observables: the bubble area, the distance from the bubble's centroid to the nearest liquid-vapor interface, and the bubble's spatial location within the simulation box, and track them over time.
During the simulation window, we cataloged a total of 30,012 bubbles.
These data show that the average lifetime of a bubble is ${1.49  \  \tau}$, with an average area of ${286  \, \sigma^2}$. 
Furthermore, our analysis of the bubble's trajectories revealed that bubble movement resembles a random walk: The center of mass of a bubble diffuses with a diffusion constant inversely proportional to the bubble radius; see Fig.~S3 in \cite{supplemental_material}. 

Fig.~\ref{fig:simulations_illustration} illustrates typical lifecycles of bubbles in a given liquid domain.
After nucleation, bubbles either remain within the domain and eventually dissolve or merge with and exit through the domain boundary.
The time sequence shown in Fig.~\ref{fig:simulations_illustration}(a) illustrates the dynamics of a bubble that dissolves internally, from nucleation to its growth (including a merging event with another bubble) and eventual disappearance.
Note that, as the bubble's area gradually diminishes to zero, it consistently remains away from the domain boundary.
Conversely, Fig.~\ref{fig:simulations_illustration}
(b) shows a bubble merging with the boundary of the liquid domain, maintaining a non-zero final area.
Notably, the vast majority of bubbles (97 \%) nucleate, grow and dissolve in the interior of the liquid domains, without any obvious interaction with the boundaries. 
Over their lifetime, these bubbles grow and shrink with considerable fluctuations in their area. 
In the following analysis, we will focus on these bubbles that do not merge with the boundary.

\begin{figure}[htb]
\includegraphics[width=\columnwidth]{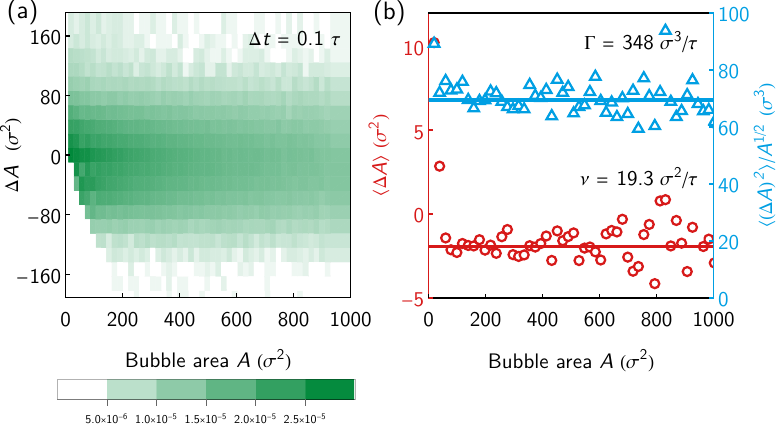}
\caption{\label{Fig2} 
Statistics of bubble area: 
(a) Probability distribution of the change of the bubble area, $\Delta A$, during a time increment ${\Delta t = 0.1 \ \tau}$, as a function of the bubble area $A$; events with ${\Delta A < - A}$ are impossible as this would lead to negative area. 
(b) First moment ($\langle \Delta A \rangle$) (circles) and second moment (${\left \langle \Delta A^2 \right \rangle}$)   (triangles), scaled by $\sqrt{A}$, of $\Delta A$ as a function of bubble area $A$. 
Least square fits (solid lines) of ${\Delta A = - \nu \Delta t}$ and ${\left \langle \Delta A^2 \right \rangle = 2 \Gamma \sqrt{A} \Delta t}$ to the data from the particle-based simulations  give values of ${\nu = 19.3 \, \sigma^2 / \tau}$ and ${\Gamma=348\, \sigma^3 / \tau}$.
}
\end{figure}

\textit{Effective Langevin dynamics for bubble area.} 
Fig.~\ref{Fig2}(a) shows the probability distribution of changes in bubble area $\Delta{A}$ over a time increment ${\Delta{t}=0.1 \, \tau}$ as a function of the initial bubble area $A$. 
To quantitatively analyze the stochastic dynamics of the bubble area, we determined its drift and diffusion, shown in Fig.~\ref{Fig2}(b). 
The drift represents the 
area change $\langle \Delta A \rangle$, averaged conditioned on a given initial area $A$, and is given by $\langle \Delta A \rangle \equiv \langle A(t)-A(t-\Delta t) \rangle = - \nu \Delta t + o(\Delta t)$. 
A least square fit to the data points gives a negative drift rate ${\nu = 19.3 \,  \sigma^2/\tau}$ that is independent of the bubble area $A$ [Fig.~\ref{Fig2}(b)]. 
Notably, the first two data points of $\langle \Delta A \rangle$ in Fig.~\ref{Fig2}(b) deviate from the average value. 
This discrepancy arises because, as shown in Fig.~\ref{Fig2}(a) and Sec.~5 of \cite{supplemental_material}, no negative increments $\Delta A$ are recorded for small bubble areas, as such increments would imply bubble sizes smaller than zero.
To quantify the stochastic dynamics of the bubble area, we determined the second moment ${\left \langle \Delta A^2 \right \rangle}$ and found that it increase with bubble area $A$ as a power law $A^{1/2}$: ${{\left \langle \Delta A^2 \right \rangle} = 2\Gamma \sqrt{A} \, \Delta t + o(\Delta t)}$ with ${\Gamma = 348 \, \sigma^3 / \tau}$. 
This can be interpreted as a configuration-dependent diffusion constant of the bubble area scaling as the bubble's perimeter. 
Taken together, our numerical results suggest that the stochastic dynamics of the bubble area $A_t$ is given by the following Langevin equation in the It\^o interpretation
\begin{equation}
   \mathrm{d} A_t
    =
    -\nu \, \mathrm{d}t 
    +
    \sqrt{2 \mathrm{\Gamma} } A_t^{1/4}  \mathrm{d}W_t \, .
    \label{eq:langevin_area}
\end{equation}
Here $\mathrm{d}W_t$ denotes statistically independent stochastic increments with a Gaussian distribution of zero mean and unit variance.  
This Langevin equation is equivalent to the Fokker-Planck equation, 
\begin{equation}
    \partial_t p(A, t)
    =
    \partial_A \big[ \nu \, p(A, t) \big]
    +
    \partial_A^2 \big[ \Gamma \sqrt{A} \, p(A, t) \big] 
    \, 
\end{equation}
with a (negative) drift term and a configuration-dependent diffusion coefficient $\Gamma \sqrt{A}$. 
The steady state solution is a `generalized Gibbs distribution' (see Sec. 8 in \cite{supplemental_material}). 
\begin{equation}
    p(A) 
    \propto 
    A^{-1/2} 
    \exp \big[ - 2\nu  \sqrt{A}/\Gamma \big] \, .
\label{eq:AreaDis}
\end{equation}
We note that the Fokker-Planck equation is distinct from Smoluchowsky's equation for a Brownian particle in a gravitational field \cite{1916Smoluchowski} by the configuration-dependence of the diffusion coefficient. In the absence of such a configuration dependence, the generalized Gibbs distribution would be reduced to the barometric height formula.
Interestingly, in terms of the effective bubble perimeter defined as ${L := 2 \sqrt{\pi A}}$, the probability distribution reads ${p(L) \sim \exp(- \lambda L)}$ with an effective line tension ${\lambda := \frac{\nu}{\Gamma \sqrt{\pi}}}$ (in units of $k_{\mathrm{B}}T$).

\begin{figure}[htb]
\includegraphics[width=\columnwidth]{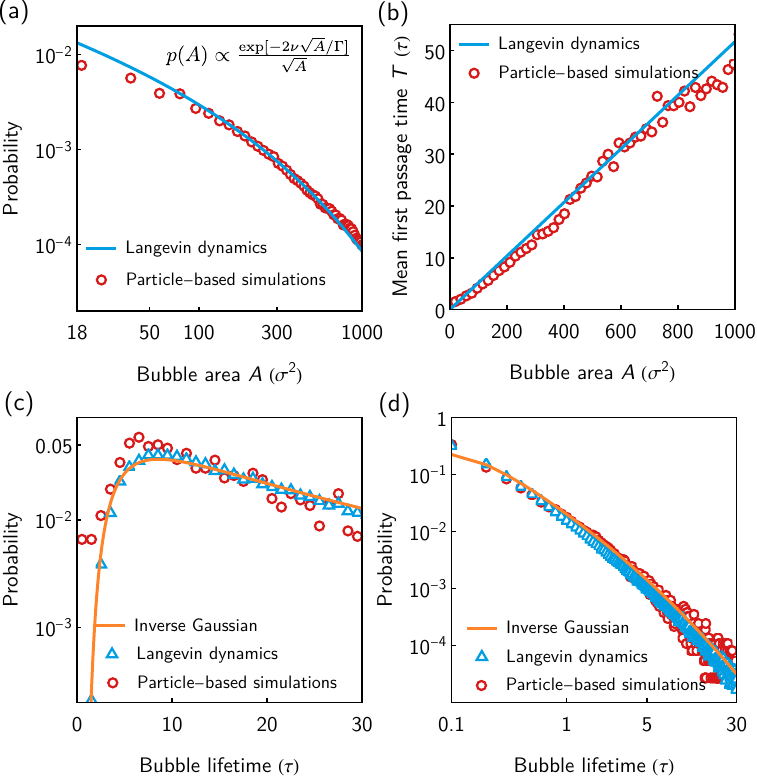}
\caption{\label{Fig3} 
Stochastic bubble dynamics:
(a) 
Stationary probability distribution $p(A)$, and 
(b) mean first-passage time $T(A)$ for the bubble area from particle-based simulations (open red circles) and the analytical solution of the Langevin dynamics (solid blue line). 
In (a), the prefactor was derived from normalization condition  ${\int_{A_\mathrm{min}}^{\infty} \mathrm{d}A \ p(A) = 1}$, with ${A_\mathrm{min}=4.375^2 \  \sigma^2}$. 
(c) The distribution of bubble lifetimes for a given initial area ${A_0 = 574 \, \sigma^2}$ and (d) of bubble lifetimes from birth to death is shown for particle-based simulations (open red circles), Langevin dynamics (open blue triangles), and the inverse Gaussian distribution (orange solid line).
In (c) and (d), the Langevin dynamics was numerically solved using the fit parameters from Fig.~\ref{Fig2}(b). 
}
\end{figure}

To test the validity of such a reduced stochastic description of the bubble-area dynamics, we compared the stationary distribution, Eq.~\eqref{eq:AreaDis}, with the results obtained from the particle-based simulations [Fig.~\ref{Fig3}(a)]. 
Except for the smallest bubble sizes, which are close to our limit of numerical resolution, there is excellent agreement.  
The results from our particle-based simulations and the effective Langevin dynamics show a significant deviation from those obtained in the continuum field theory known as Active Model B+~\cite{2018PRXABMP}, which exhibits a peak in the area distribution. 
This discrepancy suggests subtle but important differences between particle-based active models and continuum field theories. Specifically, it may be related to the gradient expansions inherent in continuum models, which may not capture certain finer details of bubble behavior observed in simulations with ABPs.

The stochastic dynamics predict that the mean first passage time for the bubble area to shrink to zero scales linearly as ${T(A) = A/\nu}$, which is quantitatively confirmed in the particle-based simulations [Fig.~\ref{Fig3}(b)]; for details on the mathematical and numerical analysis see Sec.~6 of \cite{supplemental_material}. 
This analysis of the bubble lifetime extends to its full probability distribution as a function of the initial bubble area, $P(T|A_0)$. 
While an exact solution is challenging, an approximation is obtained assuming a  Wiener process with constant effective noise amplitude: ${\mathrm{d} A_t = - \nu \, \mathrm{d} t + \sqrt{2 \Gamma} A_\text{eff}^{1/4} \, \mathrm{d} W_t}$; as discussed below we get best fits of the numerical data by choosing ${A_\text{eff}= A_0/2}$.
Then, the probability distribution for the lifetime $T$ of a bubble with initial area $A_0$ is given by the first passage time probability density to the origin 
~\cite{1915Schrödinger,1915Smoluchowski}  
\begin{equation}
    W (T | A_0)=
    \frac{A_0}{\sigma_\text{eff} \sqrt{2 \pi T^3}} \,  
    \exp 
    \left[-\frac{   (A_0 - \nu T  )^2}{2 \sigma_\text{eff}^2 T}
    \right] \, .
\end{equation} 
with an effective standard deviation ${\sigma_\text{eff} = \sqrt{2 \Gamma} A_\text{eff}^{1/4}}$; this is sometimes also referred to as the inverse Gaussian distribution~\cite{Marshall2007}.

As shown in Fig.~\ref{Fig3}(c), the inverse Gaussian solution fits both the numerical results from the particle-based simulations and the Langevin dynamics remarkably well. 
With this approximate solution of the full Langevin dynamics, we can also calculate the lifetime distribution $w(T)$ for the entire life cycle of a bubble from its birth with an area $A_\text{min}$ to its dissolution by integrating over all bubble areas, weighted with the size distribution, which leads to 
\begin{equation}
    w(T) 
    =
    \int_{A_{\rm min}}^\infty \!\!\!\!
    {\rm d}A \, p(A) 
    \int_0^T \!\!\! {\rm d}T' \; 
    W (T' | A) \, 
    W (T-T' | A) \, .
\end{equation} 
For the nucleation area $A_\text{min}$, we take the value of ${A_\text{min} = 10\  \sigma^2}$ around half of the resolution limit. 
Again, there is good agreement between these approximate results and the numerical simulations [Fig.~\ref{Fig3}(d)].

Taken together, the stochastic dynamics of the bubble area are well captured by the effective Langevin equation, Eq.~\ref{eq:langevin_area}. 
However, what about the dynamics of the bubble's shape and morphology?

\begin{figure}[!b]
\includegraphics[width=\columnwidth]{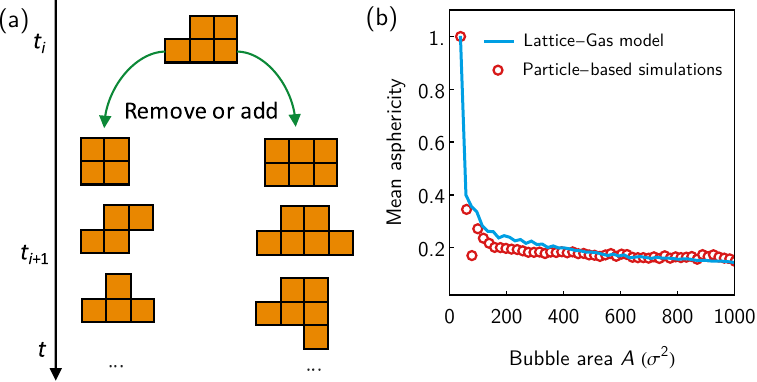}
\caption{
\label{fig:bubble_morphology} 
(a) Illustration of the lattice gas model as described in the text.  
(b) Mean asphericity from simulations of the particle-based and lattice gas model.  Simulations are made for $10^5$ bubbles with $\epsilon = 1.1$. For more information and parameters, see Sec.~10 of \cite{supplemental_material}.
}
\end{figure}

\textit{Effective stochastic dynamics of bubble shape.} 
To investigate the evolution of the bubble shape, we developed a lattice gas model that describes the dynamics of the discretized bubble shapes on the density mesh of the particle-based simulations (with mesh size ${\xi=4.375\ \sigma}$, see Sec.~2 of \cite{supplemental_material}):
Starting from a single seed, new square building blocks can be added or removed at a rate proportional to the perimeter. Fig.~\ref{fig:bubble_morphology}(a) illustrates this process, showing a bubble with an initial area of $5 \ \xi^2$ at time $t_{i}$. 
The bubble's area may increase or decrease at time $t_{i+1}$, as illustrated by the possible configurations after removing or adding an element. 
The rates for adding or removing an element anywhere on the bubble boundary (which determine the change in bubble area) are derived from the drift and diffusion parameters of the effective Langevin equation; see Sec.~10 of \cite{supplemental_material} for more details.
For the process of adding or removing an element—which determines changes in the bubble's shape—we assume that the rates at site $i$ depend on the local geometry: Sites with more occupied neighbors, $m_i$, are more likely to be  added and less likely to be removed, as described by the Boltzmann weight:
\begin{equation}
    p_i^{\text{add}}
    =
    1-p_i^{\text{remove}}
    =
    \frac{1}{Z} \, e^{\epsilon \, m_i} 
\label{eq:Boltzmann}
\end{equation}
where $\epsilon$ is the interaction scale and $Z=\sum_i e^{\epsilon m_i}$ is a normalization factor with the sum running over all possible growth sites. 

To compare our lattice gas model with bubble evolution in particle-based simulations, we analyze bubble morphology using their asphericity $\Delta$ which quantifies deviations from perfect spherical symmetry~\cite{Rudnick_1986,rudnick1987,Frey2007}; ${\Delta=0}$ indicates a perfect circle, while ${\Delta \to 1}$ indicates an elongated, rod-like object (see Sec.~9 of~\cite{supplemental_material}).
Fig.~\ref{fig:bubble_morphology}(b) shows the mean asphericity as a function of area for both particle-based simulations and the lattice gas model. For large bubbles, the two descriptions agree well, both showing a steady decrease in mean asphericity as the bubble grows larger. 
The agreement with the lattice gas suggests that the stochasticity in bubble area changes arises from random events at the bubble's boundary, such as localized growth and shrinkage.  
This aligns with the effective diffusion coefficient of bubbles decreasing as $1/\sqrt{A}$, which reflects the role of boundary fluctuations---rather than rigid-body translation---as the primary mechanism driving the center-of-mass displacement (see Fig.~S3 in \cite{supplemental_material}).

\textit{Conclusion and outlook.} 
In conclusion, we have investigated the stochastic dynamics of bubbles within liquid domains of phase-separated ABP systems, leading to two major findings. 
First, the majority of bubbles nucleate within liquid domains, grow to a certain size, and subsequently dissolve within these domains; only rarely, and particularly in smaller domains, do bubbles escape through the domain boundaries. 
Second, the stochastic dynamics of these bubbles are well-described by a Langevin equation with a noise amplitude that scales with the perimeter of the bubbles. 
This dynamics can be interpreted as an effective thermodynamic model 
${\partial_t A = - M(A) F'(A) + \sqrt{2M(A)} \, \Lambda}$, where the free energy is given by ${F(A) = \frac{2\nu}{\Gamma} \sqrt{A}}$, the mobility is ${ M(A) = \Gamma \sqrt{A}}$, and $\Lambda$ is white noise with unit variance. 
This implies that both the cost associated with creating a bubble and the fluctuations of area scale with the interface length. This suggests that 
the effective line tension, $\lambda = \nu/(\sqrt{\pi} \, \Gamma)$, as defined earlier, can indeed be interpreted as such in a consistent way, $F=2\pi \sqrt{A/\pi^2} \lambda$.

Our findings challenge the hypothesis that the Active Model B+ captures all aspects of fluctuations in the nonequilibrium steady state of ABP systems. 
Future investigations into stochastic bubble dynamics should focus on further exploring the effective Langevin dynamics and the lattice gas model we have outlined. 
Moreover, beyond the specific problem of bubble dynamics in motility-induces phase separation, our effective Langevin equation belongs to a broader class of stochastic processes with \textit{sub-demographic multiplicative noise} and area-dependent drift
\begin{equation}
    \mathrm{d} A_t
    =
    -\nu A^\beta \, \mathrm{d}t 
    +
    \sqrt{2 \mathrm{\Gamma} A_t^{\alpha} }  \, \mathrm{d}W_t 
    \,
\end{equation}
with ${0 \leq \alpha\, , \beta  \leq 1}$. This type of noise interpolates between Brownian noise (${\alpha =0 }$) and demographic noise (${\alpha = 1}$) and shows interesting properties in its own right (see Sec.~8 of \cite{supplemental_material}).
One open question this could address is the statistical properties of bubbles in three-dimensional systems of ABPs, assuming such bubbles exist \cite{cates2024activephaseseparationnew}.

\smallskip

\begin{acknowledgments}
This research was conducted within the Max Planck School Matter to Life supported by the German Federal Ministry of Education and Research (BMBF) in collaboration with the Max Planck Society.
MY thanks Kebin Sun for assistance with coding. Computations were performed on the GPU cluster at the Institute for Theoretical Physics, University of Göttingen. 
EF acknowledges financial support from the Excellence Cluster ORIGINS under Germany’s Excellence Strategy (EXC-2094-390783311) and the Chan-Zuckerberg Initiative (CZI). 
\end{acknowledgments}

\bibliography{references}
\end{document}